\documentclass[12pt]{iopart}

\usepackage{iopams}  
\usepackage{cite}

\expandafter\let\csname equation*\endcsname\relax
\expandafter\let\csname endequation*\endcsname\relax
\usepackage{amsmath}

\usepackage{mathtools}

\usepackage{mathrsfs}

\usepackage{graphicx}

\usepackage[dvipsnames]{xcolor}
\usepackage[colorlinks,linkcolor=blue,citecolor=blue]{hyperref}

\usepackage{tikz}
\usetikzlibrary{arrows.meta}

\renewcommand{\vec}[1]{\boldsymbol{#1}}
\newcommand*\diff{\mathop{}\!\mathrm{d}}

\newcommand{\vx}{\vec{x}}
\newcommand{\vn}{\vec{n}}
\newcommand{\ehat}{\hat{\vec{e}}}
\newcommand{\del}{\partial}
\newcommand{\avg}[1]{\left\langle#1\right\rangle}
\newcommand{\dx}{\diff\vx\,}
\newcommand{\om}{\Omega}
\newcommand{\po}{\partial \Omega}

\newcommand{\reals}{\mathbb{R}}

\newcommand{\hugo}[1]{\textcolor{Blue}{#1}}

\begin{document}
	
\title{Large deviations of currents in diffusions with reflective boundaries}

\author{E. Mallmin$^1$, J. du Buisson$^2$, and H. Touchette$^3$}

\address{$^1$ SUPA, School of Physics and Astronomy, University of Edinburgh, Peter Guthrie Tait Road, Edinburgh EH9 3FD, UK}
\address{$^2$ Institute of Theoretical Physics, Department of Physics, Stellenbosch University, Stellenbosch 7600, South Africa}
\address{$^3$ Department of Mathematical Sciences, Stellenbosch University, Stellenbosch 7600, South Africa}
\ead{emil.mallmin@ed.ac.uk, johan.dubuisson@gmail.com, htouchette@sun.ac.za}
\vspace{10pt}
\begin{indented}
\item[] 9 February 2021; Revised 16 June 2021
\end{indented}

\begin{abstract}
We study the large deviations of current-type observables defined for Markov diffusion processes evolving in smooth bounded regions of $\reals^d$ with reflections at the boundaries. We derive for these the correct boundary conditions that must be imposed on the spectral problem associated with the scaled cumulant generating function, which gives, by Legendre transform, the rate function characterizing the likelihood of current fluctuations. Two methods for obtaining the boundary conditions are presented, based on the diffusive limit of random walks and on the Feynman--Kac equation underlying the evolution of generating functions. Our results generalize recent works on density-type observables, and are illustrated for an $N$-particle single-file diffusion on a ring, which can be mapped to a reflected $N$-dimensional diffusion.
\end{abstract}

\begin{indented}
\item[Keywords: ] dynamical large deviations, reflected diffusions, current fluctuations, single-file diffusion
\end{indented}
\vspace{0.5cm}
Published with Open Access in \jpa:$\phantom{aaaaaaaaaaaaaaaaaaaaaaaaaaaa}$ \href{https://doi.org/10.1088/1751-8121/ac039a}{https://doi.org/10.1088/1751-8121/ac039a}


\section{Introduction}

The main property of nonequilibrium systems that distinguishes them from equilibrium systems is the existence of energy or particle currents produced by non-conservative internal or external forces, or coupling to reservoirs at different temperatures or chemical potentials \cite{zwanzig2001}. These currents and their fluctuations have been the subjects of many studies in the last decades, owing to their importance in biological transport phenomena \cite{Julicher1997,Sinitsyn2009}, and the existence of fundamental symmetries, referred to as fluctuation relations, which constrain the probability distribution of current-like quantities, such as the entropy production \cite{evans1993,gallavotti1995,lebowitz1999}.

Similarly to equilibrium systems, the fluctuations of observables, such as currents, can be studied for nonequilibrium systems using the theory of large deviations, which provides a number of general techniques for obtaining the distribution of observables in a given scaling limit (e.g., low-noise, long-time or large-volume limit) relevant to the system studied \cite{dembo1998,hollander2000,touchette2009}. Many of these techniques were successfully applied in recent years for Markov models of physical interest, including random walks, interacting particle models, such as the exclusion and zero-range models, as well as Markov diffusions described by stochastic differential equations (see \cite{derrida2007,bertini2007,harris2007,touchette2017} for useful reviews).

In this paper, we study the large deviations of Markov diffusions evolving in bounded domains of $\reals^d$ with reflection at the boundaries. Such processes are encountered in many applications, including in biology where they model the diffusion of nutrients within cells \cite{schuss2013}, and have been studied before in large deviation theory, though either in the low-noise limit \cite{dupuis1987b,sheu1998,majewski1998,ignatyuk1994} or for density-type observables defined as time integrals of the state of the process \cite{grebenkov2007,forde2015,fatalov2017,pinsky1985,pinsky1985b,pinsky1985c,budhiraja2003,Dolezal2019,Buisson2020}. Here, we consider the long-time or ergodic limit and focus on current-type observables, defined as integrals of the state and increments of the reflected diffusion. For these, we show how the large deviation  functions (viz., scaled cumulant generating function and rate function) characterizing the fluctuations of observables can be obtained from a spectral equation that must be solved with special boundary conditions, taking into account the reflection of the process at the boundaries and the fact that the observable is  current-like.

These boundary conditions generalize those found recently for density-type observables \cite{Buisson2020} and are non-trivial in that they cannot be obtained by directly applying the duality argument used in \cite{Buisson2020}. This point is explained in the next section, following a summary of large deviation theory as applied to dynamical currents defined in the long-time limit. In the sections that follow, we then derive the boundary conditions using two different methods: one based on the diffusive limit of a random walk model, which has the advantage of being physically transparent, and another, more formal method based on the Feynman--Kac equation, which underlies the spectral equation, and a local-time formalism to describe the boundary behavior.

A consequence of the boundary conditions imposed on the spectral problem is that the stationary current of the effective or driven process, introduced recently as a way to understand how large deviations are created in time \cite{evans2004,jack2010b,chetrite2013,chetrite2014}, vanishes in the normal direction of the boundaries or walls limiting the diffusion. This is a natural result given that the effective process corresponds, in an asymptotic way, to the original process conditioned on realizing a given fluctuation \cite{chetrite2014} and, therefore, should inherit the zero-current condition of the original process resulting from the reflections.

As an illustration of our results, we study the integrated particle current of a heterogeneous single-file diffusion, consisting of $N$ driven, non-identical particles diffusing on a ring without crossings \cite{Mallmin2021a}. This model can be mapped to a non-interacting diffusion model in $\reals^N$ with a reflecting wall, for which the spectral problem can be solved exactly to obtain the scaled cumulant generating function and rate function of the current, characterizing its stationary value and fluctuations. Applications to other models, including diffusions with partial or sticky reflections at boundaries, are discussed in the final section of the paper.

\section{Dynamical large deviations with and without boundaries}
\label{sec:dyn-ld}

\subsection{Formalism in unbounded domains}

We consider a $d$-dimensional Markov diffusion $\mathbf{X}(t)$ evolving in $\mathbb{R}^d$ according to the stochastic differential equation (SDE)
\begin{equation}
\label{eq:SDE}
\diff \mathbf{X}(t) = \vec{F}(\mathbf{X}(t)) + \sigma \diff \mathbf{W}(t),
\end{equation}
where the drift vector $\vec{F}$ depends on the state of the process, whereas the noise matrix $\sigma$ multiplying the Wiener noise is constant. We define the diffusion matrix $\mathsf{D} = \sigma \sigma^{\mathsf{T}}$. The process density $\rho(\vx,t) = \text{Prob}[\mathbf{X}(t) = \vx \mid \mathbf{X}(0) = \vec{x}_0]$ evolves according to the Fokker-Planck (FP) equation
\begin{equation}
\del_t \rho(\vx, t) = (\mathcal{L}^\dagger \rho)(\vx,t),
\end{equation}
expressed using the FP operator
\begin{equation}
\label{eq:Ldag}
\mathcal{L}^\dagger = - \nabla \cdot \vec{F} + \frac{1}{2} \nabla \cdot \mathsf{D} \nabla,
\end{equation}
which acts on the domain of twice-differentiable, integrable densities. We assume the existence of a unique invariant density $\rho^*$ for which $\mathcal{L}^\dagger \rho^* = 0$. Time-dependent averages of functions $f$ of the process evolve in time according to
\begin{equation}
\del_t \avg{f(\mathbf{X}(t))} = \avg{ (\mathcal{L} f)(\mathbf{X}(t))},
\end{equation} 
with the operator
\begin{equation}
\label{eq:L}
\mathcal{L} = \vec{F} \cdot \nabla + \frac{1}{2} \nabla \cdot \mathsf{D} \nabla,
\end{equation}
known as the Markov generator \cite{rogers2000}. Using the standard inner product
\begin{equation}
\avg{\rho, f} = \int_{\mathbb{R}^d} \dx \rho(\vx) f(\vx),
\end{equation}
the Markov operators $\mathcal{L}$ and $\mathcal{L}^\dagger$ are related via the duality relation
\begin{equation}
\label{eq:duality}
\avg{ \mathcal{L}^\dagger \rho, f} = \avg{\rho, \mathcal{L} f}
\end{equation} 
for arbitrary densities $\rho$ and functions $f$ for which the inner product is finite.

We are concerned in this paper with time-integrated observables $V_T$ derived from the empirical current $\mathbf{J}_T$ of the process. The empirical current at $\vx$ is the number of passes through $\vx$  (counted with sign) per unit time in a time window $[0,T]$: 
\begin{subequations}
\label{eq:j}
\begin{align}
\mathbf{J}_T(\vx) &= \frac{1}{T}\int_0^T \delta(\mathbf{X}(t) - \vx) \circ \diff\mathbf{X}(t) \\
&= \frac{1}{T} \lim\limits_{\Delta t \to 0} \sum_{i} \delta \left( \frac{\mathbf{X}(t_i+\Delta t) + \mathbf{X}(t_i)}{2} - \vec{x} \right) (\mathbf{X}(t_i+\Delta t) - \mathbf{X}(t_i)) .
\end{align}
\end{subequations}
The first line expresses a Stratonovich integral, defined as the discretization procedure shown in the second line, where $i$ indexes time-points in $[0,T]$ separated by the vanishing duration $\Delta t$. The rationale for using the Stratonovich convention is that the current is properly antisymmetric under time-reversal of trajectories and has an expectation equal in the long-time limit to the stationary current
\begin{equation}
\label{eq:J}
\vec{J}_{\vec{F},\rho^* } = \vec{F}\rho^*  - \frac{1}{2} \mathsf{D} \nabla \rho^*.
\end{equation}

From $\mathbf{J}_T(\vx)$, it is possible to define other current-like dynamical observables as
\begin{equation}
\label{eq:Gobs}
V_T = \int_{\reals^d} \vec{g}(\vx) \cdot \mathbf{J}_T (\vx)\, \dx = \frac{1}{T}\int_0^T \vec{g}(\mathbf{X}(t)) \circ \diff\mathbf{X}(t),
\end{equation}
where $\vec{g}$ is an arbitrary kernel vector field. Depending on the process and the choice of $\vec{g}$, this observable can represent, for example, a particle or energy current, or the entropy production. The probability density $P_T(v) = \text{Prob}[ V_T = v  ]$ of such observables generally satisfies a large deviation principle for large\footnote{In practice, The observation time must be larger than any relevant relaxation time scale of the process.} observation times $T$, meaning that
\begin{equation}
P_T(v) = \exp[ - T I(v) + o(T)]. 
\end{equation}
The rate function $I(v)$ characterizes the exponential decay of fluctuations $v$, that is, sustained deviations of the observable from the typical value(s) $\bar{v}$ for which $I(\bar{v}) = 0$. Dynamical large deviation theory is concerned with calculating the rate function and with  describing via an effective process the subset of process realizations that give rise to any given fluctuation. Below we outline the basic elements of dynamical large deviation theory, referring to \cite{chetrite2014} for details and derivations. 

In order to calculate the rate function, we introduce the scaled cumulant generating function (SCGF)
\begin{equation}
\lambda(k) = \lim\limits_{T\to \infty} \frac{1}{T} \ln \avg{ \exp[T k V_T] },
\end{equation}
which is related to the rate function via Legendre--Fenchel \hugo{(LF)} transform, 
\begin{equation}
\label{eq:I-LF}
I(v) = \sup_{k} \{ k v - \lambda(k)  \}.
\end{equation} 
This assumes that $I(v)$ is convex, which is guaranteed, for instance, if $\lambda(k)$ is continuously differentiable in $k$ \cite{touchette2009}. Furthermore, it can be shown (via the Feynman--Kac formula \cite{touchette2017}) that the generating function 
\begin{equation} 
\label{eq:FK-gen}
    u(\vx, t) = \avg{\exp[tk V_t]}_{\vec{x}},
\end{equation}
where the subscript $\vec{x}$ indicates the initial condition $\mathbf{X}(0) = \vec{x}$, has a semigroup structure and evolves in time according to
\begin{equation} 
\label{eq:FK-gen-evolve}
   \partial_t u(\vx,t) = \mathcal{L}_k u(\vx,t),
\end{equation}
where the `tilted' generator $\mathcal{L}_k$ is given by 
\begin{equation} \label{eq:Lk}
\mathcal{L}_k =  \vec{F} \cdot (\nabla + k \vec{g}) + \frac{1}{2} (\nabla + k \vec{g}) \cdot \mathsf{D} (\nabla + k \vec{g}).
\end{equation}
Since the semigroup equation \eqref{eq:FK-gen-evolve} is linear, we can decompose it in terms of the eigenvalues $\lambda_k^{(i)}$ and eigenfunctions $r_k^{(i)}$ of $\mathcal{L}_k$:
\begin{equation}
\label{eq:FK-spectral}
    u(\vx,t) = \sum_{i} a_i e^{\lambda_k^{(i)} t} r_k^{(i)}(\vx).
\end{equation}
Inserting this decomposition in the definition of the SCGF, we find in general that the SCGF is the dominant (Perron--Frobenius) eigenvalue of $\mathcal{L}_k$, so we can write as a shorthand
\begin{equation}
\label{eq:directspectral}
   \mathcal{L}_k r_k = \lambda(k) r_k,
\end{equation}  
$r_k$ being the eigenfunction associated with the dominant eigenvalue and SCGF $\lambda(k) = \max_i \text{Re}\; \lambda_k^{(i)}$.

These spectral elements are also used in large deviation theory to define a new Markov diffusion, called the effective or driven process, with generator \cite{chetrite2014}
\begin{equation}
\mathcal{L}_k^{\text{eff}} = r_k^{-1} \mathcal{L}_k r_k - \lambda(k), 
\end{equation}
whose invariant density is
\begin{equation}
\label{eq:rhok*}
\rho_k^*(\vx) = \ell_k(\vx) r_k(\vx),
\end{equation}
where $\ell_k$ solves the dual eigenvalue problem
\begin{equation}
\label{eq:dualspectral}
\mathcal{L}_k^\dagger \ell_k = \lambda(k) \ell_k
\end{equation}
with natural boundary conditions on $\mathbb{R}^d$. Here, $\mathcal{L}_k^\dagger$ is related to $\mathcal{L}_k$ via the duality  \eqref{eq:duality} and is given explicitly by
\begin{equation}\label{eq:Lkdag}
\mathcal{L}_k^\dagger = - (\nabla - k\vec{g}) \cdot \vec{F} + \frac{1}{2} (\nabla - k\vec{g}) \cdot \mathsf{D} (\nabla - k\vec{g}).
\end{equation}
Note that the boundary conditions on $r_k$ are defined indirectly by imposing natural boundary conditions for the density $\rho_k^*$ on $\mathbb{R}^d$.

The effective process corresponds to a process with the same diffusion matrix as the original one, but with a modified drift
\begin{equation}
\label{eq:Fk}
\vec{F}_k = \vec{F} + \mathsf{D} ( k \vec{g} + \nabla \ln r_k) .
\end{equation}
When $k$ is tuned according to 
\begin{equation} \label{eq:tuning}
k(v) = I'(v),
\end{equation}
that is, the maximizer in \eqref{eq:I-LF}, then the effective process gives $v$ as the long-time value of $V_T$. In this way, this process can be interpreted as describing (asymptotically) the set of trajectories of the original process conditioned on $V_T = v$ (see \cite{chetrite2014} for a precise statement).

An analogy with equilibrium statistical mechanics can be established by noting that the tilted generator corresponds (asymptotically) to the generator of a non-conservative process constructed by penalizing the probability of every trajectory $\mathbf{X}(t)$ over $[0,T]$ by the weight factor $\exp[T k V_T]$ as $T\to \infty$. This penalized distribution on trajectories is analogous to the `canonical ensemble', bar the missing normalization \cite{chetrite2014}. Following this analogy, the SCGF can be seen as being analogous to a free energy density, while the rate function, obtained via the LF transform, is analogous to an entropy density \cite{touchette2009}. The tuning \eqref{eq:tuning} relates $k$ to the  inverse temperature in the canonical ensemble.

\subsection{Introducing boundaries: constraints from duality}

We now consider a domain $\Omega \subset \mathbb{R}^d$ which has a smooth boundary $\del \Omega$. In the interior of the domain, the process $\mathbf{X}(t)$ is still described by the SDE \eqref{eq:SDE}. Consequently, the operators $\mathcal{L}$ and $\mathcal{L}^\dagger$ are still given by \eqref{eq:L} and \eqref{eq:Ldag}, but must be supplemented with boundary conditions on the functions $f$ and $\rho$ on which they act, in order to account for the boundary behaviour. 

These boundary conditions are related through the duality relation \eqref{eq:duality}, but with an inner product integrating over $\Omega$ rather than over all of $\mathbb{R}^d$. Starting from $\avg{ \mathcal{L} \rho, f} $ and performing repeated integration by parts to shift the derivatives from $f$ to $\rho$ (see \ref{app:duality}), one finds 
\begin{equation}
\label{eq:duality-markov}
    \avg{\rho, \mathcal{L} f} = \avg{\mathcal{L}^{\dagger} \rho, f} - \int_{\po}\dx f(\vx) \vec{J}_{\vec{F}, \rho}(\vx) \cdot \hat{\vec{n}}(\vx)  - \frac{1}{2} \int_{\po} \dx \rho(\vx) \mathsf{D} \nabla f(\vx) \cdot \hat{\vec{n}}(\vx) ,
\end{equation}
where  $\hat{\vec{n}}$ is the (inward) normal of $\del \Omega$ and the probability current $\vec{J}_{\vec{F},\rho }$ is as defined in \eqref{eq:J}.

In order for the operators $\mathcal{L}$ and $\mathcal{L}^{\dagger}$ to be well defined, independently of any particular $\rho$ or $f$, it is necessary that the surface integral terms in \eqref{eq:duality-markov} always vanish. A particular prescription that accomplishes this constitutes a boundary condition, and amounts to a restriction of the domains of the Markov operators. Note that if we put $f \equiv 1$, then the vanishing of the boundary term corresponds to conservation of probability as it represents the zero net current through the boundary.

In this paper, we are concerned with reflective boundaries. At the level of the FP equation, this means that the probability flow through the boundary vanishes at every point on the surface:
\begin{equation}
\label{eq:reflection-bc}
\vec{J}_{\vec{F}, \rho}(\vx) \cdot \hat{\vec{n}}(\vx) = 0\quad\text{for all}\quad \vx \in \del\Omega.
\end{equation} 
From \eqref{eq:duality-markov}, we note that
\eqref{eq:reflection-bc} implies that the boundary condition on $f$ is
\begin{equation}
\label{eq:f-bc}
\mathsf{D} \nabla f (\vx) \cdot \hat{\vec{n}}(\vx) = 0\quad\text{for all}\quad \vx \in \del\Omega,
\end{equation}
which is equivalent to
\begin{equation}
\nabla f (\vx) \cdot \mathsf{D}\hat{\vec{n}}(\vx) = 0\quad\text{for all}\quad \vx \in \del\Omega,
\end{equation}
since $\mathsf{D}$ is symmetric by definition. Thus, unlike the current, $\nabla f$ does not vanish in the direction normal to the surface, but in the direction $\mathsf{D}\hat{\vec{n}}$, which is called the conormal direction \cite{freidlin1985,schuss2013}.

We now turn to the large deviation elements associated with the current-like observable \eqref{eq:Gobs}. Just as in the case without boundaries, we must solve the dominant eigenvalue problem \eqref{eq:directspectral} for $\mathcal{L}_k$ and \eqref{eq:dualspectral} for its dual $\mathcal{L}_k^\dagger$, but now with boundary conditions for $r_k$ and $\ell_k$ on $\del \Omega$ in some way determined by the reflective boundary of the original process. Clearly, these boundary conditions must be consistent with \eqref{eq:reflection-bc} and \eqref{eq:f-bc} for $k=0$. In addition to this constraint, the duality relation for the eigenvectors should also hold for all $k$. 
Performing repeated integration by parts we find \eqref{app:duality2}
\begin{equation}
\label{eq:tilted-duality}
\avg{\ell_k, \mathcal{L}_k r_k}  = \avg{ \mathcal{L}^\dagger_k \ell_k, r_k} -\int_{\del \Omega}\dx \vec{J}_{\vec{F}_k, \ell_k r_k}(\vx) \cdot \hat{\vec{n}}(\vx) ,
\end{equation}
where $\vec{F}_k$ is the modified drift \eqref{eq:Fk}. Interpreting again $\ell_k r_k = \rho_k^*$ as the stationary density of an effective process with drift $\vec{F}_k$, the vanishing of the boundary term in \eqref{eq:tilted-duality} expresses the conservation of probability for the effective process. 

On physical grounds, it is reasonable to suppose that if the original process has reflective boundaries, then so does the effective process, meaning that the boundary term in \eqref{eq:tilted-duality} vanishes because
\begin{equation}\label{eq:J.n=0_eff}
    \vec{J}_{\vec{F}_k, \ell_k r_k}(\vx) \cdot \hat{\vec{n}}(\vx) = 0 \quad \textnormal{for all} \quad \vx \in \po.
\end{equation}
However, as we are about to see, this condition does not allow us to uniquely determine boundary conditions for $r_k$ and $\ell_k$ separately. Indeed, we can write, for an arbitrary constant $c$, 
\begin{subequations}
	\begin{align}\label{eq:J_Fk_lkrk-a}
	 \vec{J}_{\vec{F}_k, \ell_k r_k} &= (\vec{F} + k \mathsf{D}\vec{g} )\ell_k r_k + \frac{1}{2} \mathsf{D} \ell_k \nabla r_k -  \frac{1}{2} \mathsf{D} r_k \nabla \ell_k \\
	 & = \begin{multlined}[t]	 
	 \left[ [\vec{F} + (1-c)k \mathsf{D}\vec{g}] \ell_k -  \frac{1}{2} \mathsf{D}  \nabla \ell_k  \right] r_k   + \ell_k \left[ c k \mathsf{D}\vec{g} r_k  + \frac{1}{2} \mathsf{D}  \nabla r_k \right].
	 \end{multlined}
	\end{align}
\end{subequations}
With this identity, the boundary term in \eqref{eq:tilted-duality} can then be made to vanish by imposing the following boundary conditions:
\begin{subequations}\label{eq:c-bc}
	\begin{align}
	& \left\{ [\vec{F}(\vx) + (1-c)k \mathsf{D}\vec{g}(\vx)] \ell_k(\vx) -  \frac{1}{2} \mathsf{D}  \nabla \ell_k(\vx) \right\}  \cdot \hat{\vec{n}}(\vx) = 0 ,\\
	& \left\{ c k \mathsf{D}\vec{g} r_k  + \frac{1}{2} \mathsf{D}  \nabla r_k \right\} \cdot \hat{\vec{n}}(\vx) = 0, 
	\end{align}
\end{subequations}
for $\vx \in\del \Omega$ and any $c$. The ambiguity 
arises from the fact that any boundary term proportional to $\ell_k r_k$ that vanishes as $k\to 0$ can be split between the two conditions. In \cite{Buisson2020}, which dealt with reflected diffusions conditioned on density-like observables, this ambiguity does not arise, because the `tilting' of the generator does not produce any new boundary terms in the duality relation. An argument beyond assuming a reflective boundary for the effective process to satisfy the duality is therefore necessary to establish the correct boundary conditions.

In the next sections we provide two independent methods, free from assumptions about the boundary behavior of the effective process: one based on taking the diffusive limit of a conditioned lattice problem with boundary (Sec.~\ref{sec:difflim}), and the other based on considering local time at the boundary in the Feynman--Kac formula that defines the tilted generator (Sec.~\ref{sec:FK}). From both approaches, the boundary conditions emerge as 
\begin{subequations}\label{eq:lk-rk-bc}
	\begin{align}\label{eq:lk-bc}
	& \left\{ \vec{F}(\vx)\ell_k(\vx) -  \tfrac{1}{2} \mathsf{D} (\nabla - k \vec{g}) \ell_k(\vx) \right\}  \cdot \hat{\vec{n}}(\vx) = 0 ,\\
	& \tfrac{1}{2}\mathsf{D} (\nabla  +  k \vec{g}(\vx) )r_k(\vx)  \cdot \hat{\vec{n}}(\vx) = 0. \label{eq:rk-bc}
	\end{align}
\end{subequations}
These boundary conditions correspond to \eqref{eq:c-bc} with $c = 1/2$ and consequently prove that the effective process indeed possesses a reflective boundary, as described by \eqref{eq:J.n=0_eff}. It is striking that the boundary conditions are `tilted' in the same manner as the tilted generator itself by letting $\nabla \to \nabla + k \vec{g}$ for $r_k$ and $-\nabla \to -\nabla + k \vec{g}$ for $\ell_k$. Furthermore \eqref{eq:rk-bc} implies that, on the boundary, the normal component of the effective drift coincides with the original drift:
\begin{equation}\label{eq:Fk=F}
\vec{F}_k(\vx) \cdot \hat{\vec{n}}(\vx) = \vec{F}(\vx) \cdot \hat{\vec{n}}(\vx) \quad\text{for all}\quad \vx \in \del\Omega.
\end{equation}
This situation was shown to hold also for the large deviations of density-like observables in the presence of a reflecting boundary \cite{Buisson2020}.


Finally, we remark that in the special case of an observable satisfying
\begin{equation}\label{eq:Dgn=0}
\mathsf{D} \vec{g}(\vx) \cdot \hat{\vec{n}}(\vx) = 0 \quad \textnormal{for all} \quad \vx \in \po,
\end{equation}
the tilted boundary conditions \eqref{eq:c-bc} take the same form as the original ones, i.e., \eqref{eq:reflection-bc} and \eqref{eq:f-bc}, and are independent of $c$. The condition \eqref{eq:Dgn=0} is a necessary condition for the tilted generators to be symmetrizable.

\section{Boundary conditions from the diffusive limit}
\label{sec:difflim}

\subsection{Tilted lattice model}

A strategy to derive the correct boundary conditions on $r_k$ and $\ell_k$ is to consider the original diffusion as the limit of a jump process \cite{Gardiner2009}; that is, to set up a sequence of jump process $\mathbf{N}(t)$ on a lattice structure  $\mathscr{L}$, parametrized by the site separation $a$, together with a lattice-current observable  $\mathcal{A}_T$. The transition rates of the jump process are taken to scale with $a$ such that a diffusive limit exists, giving as $a\to 0$,  $\mathscr{L}\to \Omega$, $\mathbf{N}(t)\to \mathbf{X}(t)$, and $\mathcal{A}_T/a\to V_T$. Then also the spectral elements associated with the conditioning on $\mathcal{A}_T$ map from lattice to continuum, giving in this limit bulk and boundary equations for $\ell_k$ and $r_k$.

\begin{figure}
	\centering
\includegraphics[]{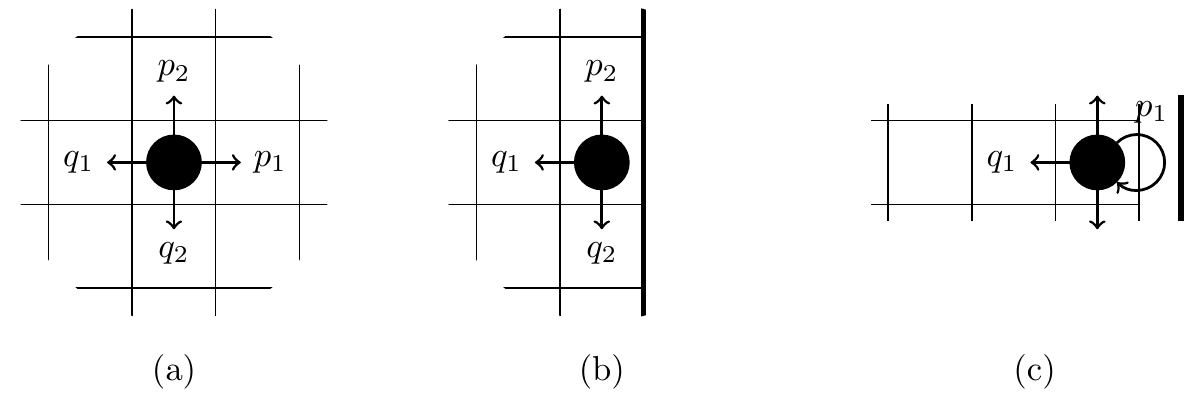}
	\caption{Illustrations of transitions available for the random walker (a) away from the boundary; (b) next to the boundary. (c) To interpret the behavior in (b) as a physical reflection, we may put the wall half a site away from the last inhabitable site, and allow hops of length $1/2$ to the wall, which get bumped off the wall $1/2$ step back, with a net translation of zero.}\label{fig:RW}
\end{figure}

For simplicity, we suppose $\mathscr{L}$ to be a cubic lattice with a planar boundary. Since the boundary $\del \Omega$ converged to is always locally planar (because it is smooth), this is not an essential limitation. The process $\mathbf{N}(t)$ evolving on $\mathscr{L}$ is then a random walk, as defined in Fig.~\ref{fig:RW}. For each $i$ labelling a spatial axis, the hopping rate is $p_i(\vec{n})$ forwards and $q_i(\vec{n})$ backwards. The hopping rate into a boundary vanishes, which can be interpreted as reflection, as explained in Fig.~\ref{fig:RW}(c).

On the lattice, we consider an observable of the form
\begin{equation}
\label{eq:Aobs}
\mathcal{A}_T= \sum_{\vn,\vn' \in \mathscr{L}} \alpha( \vn' \mid \vn ) C_T( \vn' \mid  \vn),
\end{equation}
where the empirical flow $C_T$ counts the number of transitions over the specified bond,
\begin{equation}
C_T( \vn' \mid  \vn) = \frac{1}{T} \sum_{t\in [0,T] : \mathbf{N}(t^+) \neq \mathbf{N}(t^-)} \delta_{\mathbf{N}(t^-), \vn} \delta_{\mathbf{N}(t^+),\vn'},
\end{equation}
and the function $\alpha( \vn' \mid \vn )$ represents a weight associated with the empirical flow.

Because we seek to map $\mathcal{A}_T$ onto the diffusion observable $V_T$, we choose an antisymmetric $\alpha$:
\begin{equation}
\alpha (\vn'\mid\vn ) =  - \alpha( \vn \mid \vn').
\end{equation}
We can then write \eqref{eq:Aobs} as
\begin{equation}
\label{eq:jumpcontraction}
\mathcal{A}_T = \sum_{\vn} \sum_i \alpha( \vn + \ehat_i \mid \vn ) J_T( \vn + \ehat_i \mid  \vn)
\end{equation}
where $\ehat_i$ is the single-site translation vector for axis $i$ and $J_T( \vn'\mid  \vn)$ is the empirical lattice current, defined in terms of the empirical flow as
\begin{equation}
J_T( \vn' \mid  \vn)  = C_T( \vn' \mid  \vn) - C_T( \vn \mid  \vn').
\end{equation}
The contraction of the current in \eqref{eq:jumpcontraction} giving $\mathcal{A}_T$ is the jump process analog of \eqref{eq:Gobs} for the diffusion, with $\alpha$ playing the role of $\vec{g}$.

As for diffusions, the SCGF of $\mathcal{A}_T$ corresponds to a dominant eigenvalue, this time of a $|\mathscr{L}| \times |\mathscr{L}|$ matrix $\mathbb{L}_s$ with elements \cite{chetrite2014}
\begin{equation}
\mathbb{L}_s(\vn', \vn) = W (\vn \mid \vn')  e^{s  \alpha(\vn \mid \vn')} - \delta_{\vn,\vn'} \sum_{\vn''} W(\vn''\mid\vn).
\end{equation}
The non-zero transition rates in this expression are
\begin{subequations}
\begin{align}
W(\vn + \ehat_i \mid \vn) = p_i(\vn),\quad \vn+\ehat_i \in \mathscr{L},\\
W(\vn - \ehat_i \mid \vn) = q_i(\vn),\quad \vn-\ehat_i \in \mathscr{L}.
\end{align}
\end{subequations}

\subsection{Diffusive limit of the observable}

The diffusive limit relating the master equation of $\mathbf{N}(t)$ to the FP equation of $\mathbf{X}(t)$  is defined by the following scaling relations \cite{Gardiner2009}:
\begin{subequations}
\begin{align}
F_i (\vx) =  \lim_{a\to 0} a( p_i(\vn) - q_i(\vn))\\
\sigma^2_i =    \lim_{a\to 0} a^2( p_i(\vn) + q_i(\vn)),
\end{align}
\end{subequations}
where points in $\Omega$ relate to points in $\mathscr{L}$ as  $\vx = a \vn$. Equivalently, we may state
\begin{subequations}
\begin{align}
p_i(\vn) = \frac{\sigma^2}{2a^2} + \frac{F_i(\vx)}{2a} + \mathcal{O}(1),\\
q_i(\vn) = \frac{\sigma^2}{2a^2} - \frac{F_i(\vx)}{2a} + \mathcal{O}(1). \label{eq:qi}
\end{align}
\end{subequations}

To show that $\mathcal{A}_T/a$ converges to $V_T$ in this limit, we first use the fact that $\alpha$ is antisymmetric to write
\begin{subequations}
\begin{align}
\alpha(\vn \pm \ehat_i \mid \vn) / a^d &= G(\vx \pm a \ehat_i) - G(\vx) \\
&= \pm a  g_i(\vx) + \frac{a^2}{2} \del_{x_i} g_i(\vx) + \mathcal{O}(a^3) 
\end{align}
\end{subequations}
where $G$ is an arbitrary smooth function independent of $a$ such that $g_i = \del_{x_i} G$.

Next, we note that the lattice empirical current over the $ \vn \to \vn + \ehat_i$ bond is
\begin{subequations}
	\begin{align}
	J_T( \vn + \ehat_i \mid \vn) &= \frac{1}{T} \sum_t \left[ \delta_{\mathbf{N}(t^-), \vn} \delta_{\mathbf{N}(t^+),\vn + \ehat_i} - \delta_{\mathbf{N}(t^-),\vn +\ehat_i} \delta_{\mathbf{N}(t^+),\vn} \right]\\
	&=  \frac{1}{T} \sum_t \left[ \delta_{\mathbf{N}(t^-),\vn} \delta_{\mathbf{N}(t^+),\vn+\ehat_i} + \delta_{\mathbf{N}(t^-),\vn+\ehat_i} \delta_{\mathbf{N}(t^+),\vn} \right] (N_i(t^+)- N_i(t^-)), \label{eq:Jlatt-b}
	\end{align}
\end{subequations}
where the second line follows because $N_i(t^+)$ and $N_i(t^-)$ differ by precisely one step. We now discretize time into points $t_j$ narrowly separated by intervals $\Delta t(a)$, such that the jump process makes at most one jump in each interval for any value of the site separation $a$. For any such trajectory,
\begin{equation}
\delta_{\mathbf{N}(t_j),n} \delta_{\mathbf{N}(t_j+\Delta t),\vn + \ehat_i} + \delta_{\mathbf{N}(t_j),\vn + \ehat_i} \delta_{\mathbf{N}(t_j+\Delta t),\vn} = \delta_{\frac{\mathbf{N}(t_j)+\mathbf{N}(t_j+\Delta t)}{2}, \vn + \frac{1}{2} \ehat_i},
\end{equation}
which is seen from the fact that both sides are symmetric under exchange of $\mathbf{N}(t^-)$ and $\mathbf{N}(t^+)$. In the diffusive limit we replace $\mathbf{N}(t) = \mathbf{X}(t)/a$, $\delta_{\vn,\vn'} = \delta(\vx/a-\vx'/a)$, and thus
\begin{subequations}
	\begin{align}
	J_T( \vn + \ehat_i \mid \vn) &= \frac{1}{T} \sum_j \delta_{\frac{\mathbf{N}(t_j)+ \mathbf{N}(t_j+\Delta t)}{2}, \vn + \frac{1}{2} \ehat_i} (N_i(t_j+\Delta t) - N_i (t_j))\\
	&= \frac{1}{T} \sum_j  \delta\left( \frac{\mathbf{X}(t_j+\Delta t) + \mathbf{X}(t_j)}{2} - x \right) (X_i(t_j + \Delta t) - X_i(t_j))\\
	&\hspace{-1.2mm}\overset{a \to 0}{=} 
	\ehat_i \cdot \mathbf{J}_T(\vx),
	\end{align}
\end{subequations}
as defined in \eqref{eq:j}. Hence
\begin{subequations}
\begin{align}
\mathcal{A}_T / a &= \sum_{\vn} \sum_i  \alpha(\vn + \ehat_i \mid \vn) J(\vn + \ehat_i \mid \vn) / a \\
&= \int \frac{\diff\vx}{a^{d}} \sum_i a^{d+1} g_i(\vx) \ehat_i \cdot   \mathbf{J}_T(\vx) / a + \mathcal{O}(a)\\
& \hspace{-1.2mm}\overset{a \to 0}{=}  V_T.
\end{align}
\end{subequations}

\subsection{Diffusive limit of the spectral elements}

We now derive the diffusive limit of the spectral elements, assuming the following diffusive scaling between the lattice and continuum elements:
\begin{subequations}
\begin{align}
& s = a^d k\\
&\Lambda_s = \lambda(k) + \mathcal{O}(a)\\
&L_s(\vec{n}) = a^d \ell_k(\vx) + \mathcal{O}(a^{d+1}) \\
&R_s(\vec{n}) = a^d r_k(\vx) + \mathcal{O}(a^{d+1}) 
\end{align}
\end{subequations}
To begin, we consider the limit of the right eigenvalue equation,
\begin{equation}
\Lambda_s R_s(\vn) = \sum_{\vn'} \mathbb{L}_s(\vn,\vn') R_s(\vn'). 
\end{equation}
For $\vn$ away from the boundary sites,
\begin{align} 
	\Lambda_s R_s(\vn) & =  \label{eq:Rk-free}
	\begin{multlined}[t] \sum_i \Big[
	p_i(\vn) e^{ s\,\alpha(\vn+\ehat_i \mid \vn)} R_s(\vn + \ehat_i)  
	+q_i(\vn) e^{ s\, \alpha(\vn - \ehat_i \mid \vn)} R_s(\vn - \ehat_i)  \\ - (p_i + q_i)(\vn)R_s(\vn) \Big].	
	\end{multlined}
\end{align}
Up to relevant orders in $a$, and suppressing the function arguments $\vx$ and $\vn$, 
\begin{subequations}\label{eq:rk-difflim-free}
\begin{align}
\lambda(k) r_k &=  \begin{multlined}[t]
  \sum_i \bigg[  p_i \left( 1 + k a g_i + \frac{1}{2}k a^2 \del_{x_i} g_i + k^2 a^2 g_i^2 \right)\left( r_k + a \del_{x_i} r_k + \frac{1}{2}\del_{x_i}^2 r_k  \right)\\
  + q_i \left( 1 - k a g_i + \frac{1}{2}k^2a^2 \del_{x_i} g_i+ k^2 a^2 g_i^2 \right)\left( r_k - a \del_{x_i} r_k+ \frac{1}{2}\del_{x_i}^2 r_k  \right) \\
  - (p_i + q_i) r_k \bigg]
\end{multlined}\\
&= \begin{multlined}[t]\sum_{i} \bigg[  a(p_i - q_i) \left( \del_{x_i} r_k + k g_i r_k \right) + \frac{a^2}{2}(p_i + q_i) \times \\\left( \del_{x_i}^2 r_k + r_k \del_{x_i} g_i  + 2k g_i \del_{x_i} r_k + k^2 g_i^2 r_k \right) \bigg]
\end{multlined}\\
&= \sum_{i} \left\{ F_i (\del_{x_i} + k g_i)r_k + (\del_{x_i} + k g_i) \frac{\sigma_i^2}{2} (\del_{x_i} + k g_i) r_k \right\} \\
& = \mathcal{L}_k r_k,
\end{align}
\end{subequations}
with $\mathcal{L}_k$ as in \eqref{eq:Lk}. This recovers the spectral equation for $r_k$ in the bulk.

Now let us take $\vn$ to be a boundary site as in Fig.~\ref{fig:RW}(b). Then 
\begin{align} 
	\Lambda_s R_s(\vn) & = \begin{multlined}[t]
	q_1(\vn) e^{s\,\alpha(\vn-\ehat_1 \mid \vn) }R_s(\vn-\ehat_1) - q_1(\vn) R_s(\vn) \\ 
	+ \sum_{i>1} \Big[
	p_i(\vn) e^{ s\,\alpha(\vn+\ehat_i \mid \vn)} R_s(\vn + \ehat_i)  
	+q_i(\vn) e^{ s\, \alpha(\vn - \ehat_i \mid \vn)} R_s(\vn - \ehat_i)  \\ - (p_i + q_i)(\vn)R_s(\vn) \Big].
	\end{multlined}	
\end{align}
Thus, including all relevant orders,
\begin{equation}\label{eq:rk-difflim-boundary}
\lambda(k) r_k = q_1 \left( 1  - k a g_1 \right)\left( r_k - a \del_{x_1} r_k \right) - q_1 r_k + \mathcal{O}(1),
\end{equation}
where we have used \eqref{eq:rk-difflim-free} to neglect the sum on the ${i>1}$ terms. In fact, the right-hand side of \eqref{eq:rk-difflim-boundary} is $\mathcal{O}(1/a)$. Substituting $q_i$ with \eqref{eq:qi}, multiplying both sides by $a$, and taking $a \to 0$, we then arrive at
\begin{equation}
0 = \frac{1}{2} \sigma_1^2 \left( \del_{x_1} r_k + g_1 r_k \right),
\end{equation}
which generalizes, including all other components, to
\begin{equation}
\mathsf{D} ( \nabla + k \vec{g}) r_k \cdot \hat{\vec{n}} = 0.
\end{equation}
Thus, we find the boundary condition \eqref{eq:rk-bc}, corresponding to \eqref{eq:c-bc} with $c = 1/2$.

Now that the boundary condition on $r_k$ has been established, the boundary condition \eqref{eq:lk-bc} for $\ell_k$ follows uniquely from duality. One may also verify that this boundary condition follows from the diffusive limit of the left eigenvalue equation, in a calculation analogous to that of $r_k$. Furthermore, the duality relation  \eqref{eq:tilted-duality} is the result of applying the diffusive limit to the trivial identity
\begin{equation}
 \sum_{\vn, \vn'} L_s(\vn) \mathbb{L}_s(\vn,\vn') R_s(\vn') = \sum_{\vn, \vn'}  R_s(\vn) \mathbb{L}_s^\top (\vn,\vn') L_s(\vn').  
\end{equation} 

\section{Boundary conditions from Feynman-Kac expectation}\label{sec:FK}

We provide in this section an alternative derivation of the boundary condition \eqref{eq:rk-bc} on $r_k$, proceeding directly from the generating function $u(\vx, t)$, as defined in \eqref{eq:FK-gen},  which, from its spectral decomposition \eqref{eq:FK-spectral}, shares the boundary conditions placed on the eigenfunctions of $\mathcal{L}_k$. To account for the reflection upon reaching the boundary $\del \Omega$, we employ a formulation of reflected SDEs, introduced by Skorokhod \cite{skorokhod1961stochastic} and Tanaka \cite{tanaka1979}, based on the following modified SDE:
\begin{equation} 
\label{eq:localtimesde}
   \diff \mathbf{X}(t) = \vec{F}(\mathbf{X}(t))dt + \sigma \diff \mathbf{W}(t) + \hat{\vec{\gamma}}(\mathbf{X}(t)) \diff L(t).
\end{equation}
The first two terms on the right-hand side describe the evolution of $\mathbf{X}(t)$ inside the domain $\om$, whereas the last term pushes the process inside $\om$ in the direction of the unit vector $\hat{\vec{\gamma}}(\vx)$ in the event that the process reaches $\vx \in \po$. The extra random process $L(t)$ accounting for the reflection is called the local time, since it is incremented only when the process reaches $\del\om$, and is known \cite{grebenkov2019probability} to be such that
\begin{equation}
    \avg{\diff L(t)}_{\mathbf{X}(t) = \vx}  = \mathcal{O}(\sqrt{dt}).
\end{equation}
Therefore, the increment $\diff \mathbf{X}(t)$ on $\vx \in \po$ satisfies
\begin{equation} \label{eq:update-ref}
    \avg{\diff \mathbf{X}(t)}_{\mathbf{X}(t) = \vx} = \vec{F}(\vx) dt + \sigma \avg{\diff \mathbf{W}(t)} + \hat{\vec{\gamma}}(\vx) \avg{\diff L(t)}_{\mathbf{X}(t) = \vx} = \hat{\vec{\gamma}}(\vx) \epsilon + \mathcal{O}(\epsilon^2) 
\end{equation}
where we have used the fact that $\avg{\diff \mathbf{W}(t)} = \vec{0}$ for all $t$, and where $\epsilon = \mathcal{O}(\sqrt{dt})$ such that $dt(\epsilon) = \mathcal{O}(\epsilon^2)$. Here, we take $\hat{\vec{\gamma}}$ to be in the conormal direction, that is,
\begin{equation} \label{eq:co-normal}
    \hat{\vec{\gamma}}(\vx) = \frac{\mathsf{D} \hat{\vec{n}}(\vx)}{\lvert\mathsf{D} \hat{\vec{n}}(\vx)\rvert}.
\end{equation}
This choice is necessary (see \cite[Thm.~2.6.1]{schuss2015brownian}) to preserve the zero-current condition \eqref{eq:reflection-bc} associated with reflections.

Our goal now is to understand the effect of the boundary dynamics on the generating function $u(\vx,t)$ of the observable $V_T$, as defined in \eqref{eq:FK-gen}. To this end, we consider a point $\vx \in \po$ and write the generating function as
\begin{equation}
    u(\vx,t) = \avg{\exp \left[k \int_0^{dt(\epsilon)} \vec{g}(\mathbf{X}(s))\circ \diff\mathbf{X}(s) +\int_{dt(\epsilon)}^t \vec{g}(\mathbf{X}(s))\circ \diff\mathbf{X}(s) \right]}_{\vx},
\end{equation}
having isolated in the first integral the contribution from the reflection, which takes place over the infinitesimal time $dt(\epsilon)$. Using the Stratonovich discretization, as in \eqref{eq:j}, we have 
\begin{equation}
    \exp\left[k \int_0^{dt(\epsilon)} \vec{g}(\mathbf{X}(s))\circ \diff\mathbf{X}(s)\right] = \exp\left[k \vec{g}\left(\vx + \frac{\diff \mathbf{X}(0)}{2} \right)\cdot \diff \mathbf{X}(0)\right],
\end{equation}
so that
\begin{equation}
    u(\vx,t) = \avg{\exp\left[k \vec{g}\left(\vx + \frac{\diff \mathbf{X}(0)}{2} \right)\cdot \diff \mathbf{X}(0)\right] \exp\left[\int_{dt(\epsilon)}^t \vec{g}(\mathbf{X}(s))\circ \diff\mathbf{X}(s) \right]}_{\vx}.
\end{equation}
The expression of the expectation can be written explicitly as
\begin{align}
\label{eq:u-general1}
    u(\vx,t) = \begin{multlined}[t] \int \diff (\hat{\vec{\xi}}\delta)\,  p\big(\diff\mathbf{X}(0) = \hat{\vec{\xi}}\delta \big| \mathbf{X}(0) = \vx \big) \\ e^{k\vec{g}\left(x + \frac{\hat{\vec{\xi}}\delta}{2} \right)\cdot\hat{\vec{\xi}}\delta} \avg{\exp\left[\int_{dt(\epsilon)}^t \vec{g}(\mathbf{X}(s))\circ \diff\mathbf{X}(s) \right]}_{\mathbf{X}(dt) = \vx + \hat{\vec{\xi}}\delta} \end{multlined}
\end{align}
{using the conditional probability density of the first increment $d\mathbf{X}(0)$ from $\mathbf{X}(0)= \vec{x}\in\del\Omega,$ which includes all the information about the reflections on the boundary. Using the definition of the generating function for the last factor in the integral, we then obtain
\begin{equation}
u(\vx,t) = \int \diff (\hat{\vec{\xi}}\delta)\,  p\big(\diff\mathbf{X}(0) = \hat{\vec{\xi}}\delta \big| \mathbf{X}(0) = \vx \big) e^{k\vec{g}\left(x + \frac{\hat{\vec{\xi}}\delta}{2} \right)\cdot\hat{\vec{\xi}}\delta} u(\vx + \hat{\vec{\xi}}\delta, \, t - dt(\epsilon)).
\end{equation}

%

At this point, we perform Taylor expansions in both space and time, starting with the one in space, which gives to first order in $\delta$:
\begin{align} \label{eq:u-taylor}
    u(\vx,t) = \begin{multlined}[t] \int \diff (\hat{\vec{\xi}}\delta)\,  p\big(\diff\mathbf{X}(0) = \hat{\vec{\xi}}\delta \big| \mathbf{X}(0) = \vx \big) \bigg[u(\vx, t - dt(\epsilon)) \\ + \nabla u(\vx, t - dt(\epsilon)) \cdot \hat{\vec{\xi}} \delta    + k u(\vx, t - dt(\epsilon))\vec{g}(\vx)\cdot \hat{\vec{\xi}} \delta + \mathcal{O}(\delta^2)\bigg].
    \end{multlined}
\end{align}
This becomes
\begin{align} 
\label{u-post-taylor}
    u(\vx,t) = \begin{multlined}[t]
     u(\vx, t - dt(\epsilon)) + \nabla u(\vx, t - dt(\epsilon)) \cdot \avg{\diff \mathbf{X}(0)}_{\mathbf{X}(0) = \vx} \\ + k u(\vx, t - dt(\epsilon)) \vec{g}(\vx)\cdot \avg{\diff \mathbf{X}(0)}_{\mathbf{X}(0) = \vx}   + \avg{\lvert\diff \mathbf{X}(0)\rvert^2}_{\mathbf{X}(0) = \vx},
	\end{multlined}
\end{align}
given that
\begin{equation}
     \int \diff (\hat{\vec{\xi}}\delta)\,  p\big(\diff\mathbf{X}(0) = \hat{\vec{\xi}}\delta \big| \mathbf{X}(0) = \vx \big) = 1
\end{equation}
and 
\begin{equation}
    \int \diff (\hat{\vec{\xi}}\delta)\,  p\big(\diff\mathbf{X}(0) = \hat{\vec{\xi}}\delta \big| \mathbf{X}(0) = \vx \big) \hat{\vec{\xi}}{\delta} = \avg{\diff \mathbf{X}(0)}_{\mathbf{X}(0) = \vx} ,
\end{equation}
and noting that $\delta = \lvert \diff \mathbf{X}(0)\rvert$. Moreover, since $\vx \in \po$, we have
\begin{equation}
    \avg{\diff \mathbf{X}(0)}_{\mathbf{X}(0) = \vx} = \frac{\mathsf{D} \hat{\vec{n}}(\vx)}{\lvert\mathsf{D} \hat{\vec{n}}(\vx)\rvert} \epsilon + \mathcal{O}(\epsilon^2)
\end{equation}
from \eqref{eq:update-ref} and \eqref{eq:co-normal}, yielding
\begin{align} 
    u(\vx,t) = \begin{multlined}[t]
        u(\vx, t - dt(\epsilon)) + \nabla u(\vx, t - dt(\epsilon)) \cdot \frac{\mathsf{D} \hat{\vec{n}}(\vx)}{\lvert\mathsf{D} \hat{\vec{n}}(\vx)\rvert} \epsilon  \\ + k u(\vx, t - dt(\epsilon)) \vec{g}(\vx)\cdot \frac{\mathsf{D} \hat{\vec{n}}(\vx)}{\lvert\mathsf{D} \hat{\vec{n}}(\vx)\rvert} \epsilon  + \mathcal{O}(\epsilon^2).
        \end{multlined}
\end{align}

Considering now the Taylor expansion in time, we have
\begin{equation}
    u(\vx,t - dt(\epsilon)) = u(\vx,t) - \mathcal{L}_k dt(\epsilon) u(\vx,t) + \mathcal{O}(dt(\epsilon)^2) = u(\vx,t) + \mathcal{O}(\epsilon^2),
\end{equation}
using the fact that $dt(\epsilon) = \mathcal{O}(\epsilon^2)$. Therefore, we find 
\begin{equation}
\label{eq:u-intermediate}
    u(\vx,t) = u(\vx,t) + \nabla u(\vx, t) \cdot \frac{\mathsf{D} \hat{\vec{n}}(\vx)}{\lvert\mathsf{D} \hat{\vec{n}}(\vx)\rvert} \epsilon  + k u(\vx, t ) \vec{g}(\vx)\cdot \frac{\mathsf{D} \hat{\vec{n}}(\vx)}{\lvert\mathsf{D} \hat{\vec{n}}(\vx)\rvert} \epsilon  + \mathcal{O}(\epsilon^2)\end{equation}
that is,
\begin{equation}
    \nabla u(\vx,t) \cdot \frac{\mathsf{D} \hat{\vec{n}}(\vx)}{\lvert\mathsf{D} \hat{\vec{n}}(\vx)\rvert} = -k u(\vx,t) \vec{g}(\vx)\cdot \frac{\mathsf{D} \hat{\vec{n}}(\vx)}{\lvert\mathsf{D} \hat{\vec{n}}(\vx)\rvert}
\end{equation}
or 
\begin{equation} \label{eq:generating-boundary}
    \mathsf{D} \nabla u(\vx,t) \cdot \hat{\vec{n}} =-  k u(\vx,t) \mathsf{D}\vec{g}(\vx)\cdot \hat{\vec{n}}(\vx),
\end{equation}
using the fact that $\mathsf{D}$ is symmetric. This is the boundary condition satisfied by the generating function for all $\vx \in \po$. The eigenfunctions $r_k^{(i)}(\vx)$ of the tilted generator $\mathcal{L}_k$ share the same boundary condition, so we have in the end
\begin{equation}
    \mathsf{D}\nabla r_k(\vx) \cdot \hat{\vec{n}}(\vx) = - k r_k(\vx) \mathsf{D}\vec{g}(\vx)\cdot \hat{\vec{n}}(\vx) \quad \textnormal{for all} \quad \vx \in \po,
\end{equation}
which reproduces the boundary condition \eqref{eq:rk-bc}, obtained also from the diffusive limit. The boundary condition on $\ell_k$ can then be found in the usual manner via the duality relation \eqref{eq:tilted-duality}, thereby recovering \eqref{eq:lk-bc}.

The same calculation can be performed, in principle, for an arbitrary reflection direction $\hat{\vec{\gamma}}(\vx)$, in which case the boundary condition becomes 
\begin{equation}
    \nabla u(\vx, t) \cdot \hat{\vec{\gamma}}(\vx)   = - k u(\vx, t ) \vec{g}(\vx)\cdot \hat{\vec{\gamma}}(\vx).
\end{equation}
Comparing with the duality relation \eqref{eq:duality-markov} for $k = 0$ then shows that we only obtain the zero-current condition at the boundary when the chosen direction for reflection is the conormal direction, as mentioned before. It is also interesting to note that, if we repeat the calculation for a density-type observable of the form
\begin{equation}
A_T = \frac{1}{T} \int_0^T f(\mathbf{X}(s)) \diff s,
\end{equation}
then the boundary condition is
\begin{equation}
    \mathsf{D}\nabla r_k(\vx) \cdot \hat{\vec{n}}(\vx) = 0,
\end{equation}
which is the result obtained in \cite{Buisson2020} using only the duality relation.

\section{Exactly solvable example: Heterogeneous single-file diffusion on a ring}

\begin{figure}[t]
	\centering
	\includegraphics[]{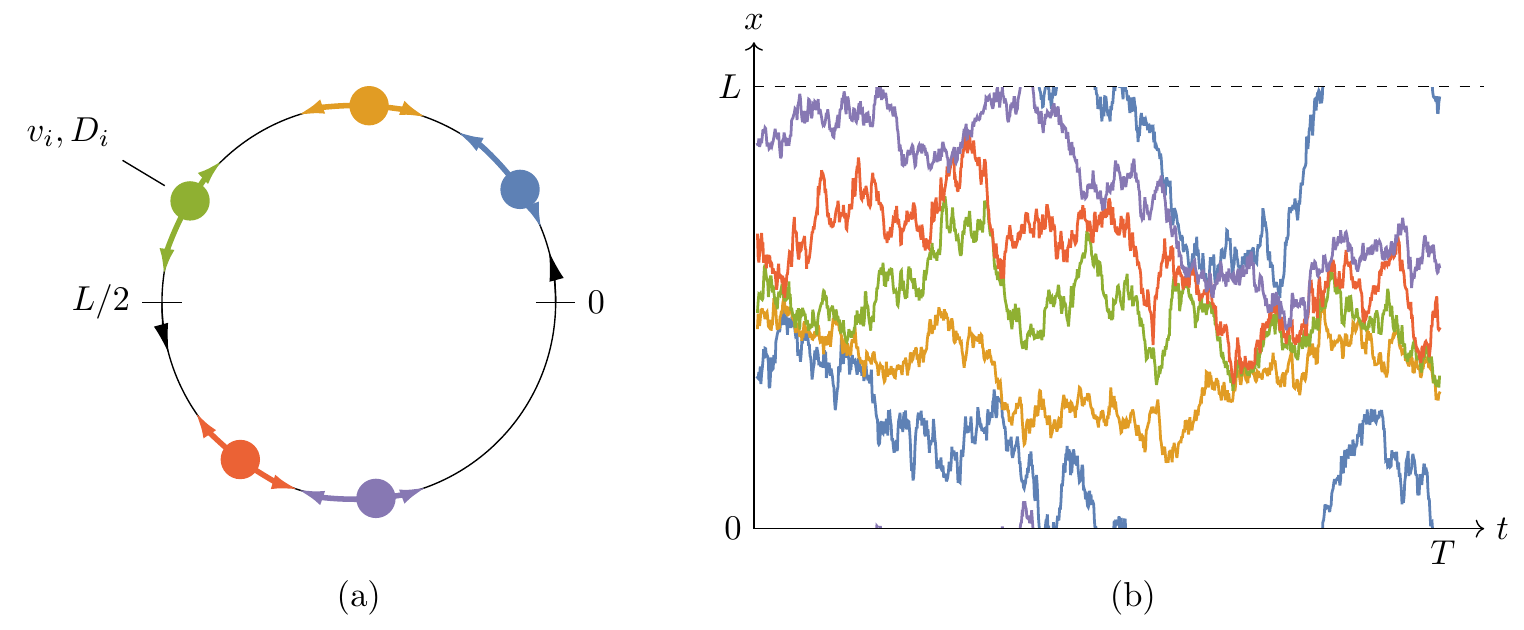}
	\caption{(a) Illustration of heterogeneous single-file diffusion on a ring. (b) Space-time plot of a typical realization. Due to hardcore exclusion, i.e.\ reflection, the particles' paths do not cross.}\label{fig:SFD}
\end{figure}

We illustrate the formalism and results developed in the previous sections for a single-file diffusion model, recently solved for its steady state \cite{Mallmin2021a} similarly to earlier lattice models \cite{Evans1996,Krug1996}. The model consists of $N$ distinct point-particles moving on a ring $S$ of circumference $L$, as illustrated in Fig.~\ref{fig:SFD}. Each particle $i$ has a constant intrinsic velocity $v_i$ and a diffusivity $D_i = \sigma_i^2$ arising from white noise of strength $\sigma_i$. The particles interact through volume exclusion: if one particle attempts to overtake another, that move is reflected. Collecting the (stochastic) particle positions $X_i(t)$ taking values in $S$ into a vector $\mathbf{X}(t)$ on a domain $\Omega \subset S^N$, we obtain an $N$-dimensional diffusion of the form \eqref{eq:SDE} where the drift $\vec{F}$ is the collection $\vec{v} = (v_1,\ldots,v_N)^\top$ of intrinsic velocities, and $\mathsf{D}$ is the diagonal matrix $\text{diag}\{D_1, \ldots, D_N \}$. 

The boundary $\del \Omega$ of the process consists of those configurations for which two (or more) particles are immediately adjacent. The hardcore exclusion rule translates into the reflective boundary condition \eqref{eq:reflection-bc}. For two particles, for instance, the boundary consists of $X_1 = X_2$, which in the space $S^2 \simeq [0,L)\times[0,L)$ is a diagonal with normal $\hat{\vec{n}} = (\hat{\vec{e}}_2 - \hat{\vec{e}}_1)/\sqrt{2} =  (-1,+1)/\sqrt{2}$. Generalising to $N$ particles, we then find that the boundary conditions are
\begin{equation}
\hat{\vec{e}}_i \cdot \vec{J}_{\vec{v},\rho}(\vx,t)|_{x_i = x_j} = \hat{\vec{e}}_j \cdot \vec{J}_{\vec{v},\rho}(\vx,t)|_{x_i = x_j} .
\end{equation}
The periodicity of the ring is implemented by the condition
\begin{equation}\label{eq:periodicity}
\rho(\vx,t) = \rho(\vx + L\vec{1}, t), 
\end{equation}
for the density, where $\vec{1}$ is the vector of ones, so that $ L\vec{1}$ is the translation vector moving all particles simultaneously by one period $L$.

It was shown in \cite{Mallmin2021a} that the invariant density of this process is
\begin{equation}\label{eq:rho-sfd}
\rho^*(\vx) \propto \exp[ \vec{k} \cdot \vx  ]. 
\end{equation}
This result assumes that $\vx \in [0,L)^N$ and that the ordering of particles is consistent with that of the initial condition, clearly conserved by the dynamics. Without loss of generality, we assume $\ldots x_{i-1} < x_i < x_{i+1} < \ldots $ (modulo $L$). The vector $\vec{k}$ has elements
\begin{equation}
k_i = \frac{v_i}{D_i} - \frac{\tilde{v}}{D_i},
\end{equation}
where $\tilde{v}$ satisfies
\begin{equation}\label{eq:v-D-tilde}
\frac{\tilde{v}}{\widetilde{D} } = \sum_{i=1}^N \frac{v_i}{D_i},\quad \text{and}\quad \frac{1}{\widetilde{D}} = \sum_{i=1}^N \frac{1}{D_i}.
\end{equation}
It is clear from the geometrical constraints on the particles' motion that they must have a common net velocity in the long-time limit, corresponding in fact to $\tilde{v}$. If follows from \eqref{eq:rho-sfd} that 
\begin{equation}\label{eq:J=v rho}
\hat{\vec{e}}_i \cdot  \vec{J}_{\vec{v},\rho^*}(\vx) = \tilde{v} \rho^*(\vx),
\end{equation}
independent of $i$, which confirms the interpretation of $\tilde{v}$. Note that $\vec{k} \cdot \vec{1} = 0$, which ensures the periodicity \eqref{eq:periodicity}.

It is natural to consider as a current-like observable the empirical velocity of particle $i$, given by the $i$th component of the empirical current $\mathbf{J}_T$ integrated over $\Omega$. However, since all particles must have the same net velocity for long averaging periods, all observables of the form \eqref{eq:Gobs} with $\vec{g}$ a constant vector whose components sum to one ($\vec{1} \cdot \vec{g} = 1$) should have the same large deviations. To validate this claim, we keep $\vec{g}$ arbitrary apart from these constraints, and thus consider the current observable
\begin{equation}
V_T = \vec{g} \cdot \int_\Omega \mathbf{J}_T(\vx)\, \dx.
\end{equation}   
The empirical velocity of particle $i$ is obtained by choosing $g_j = \delta_{ij}$.

To find the dominant eigenvalue $\lambda(k)$ and eigenvector $r_k$ related to this observable, we consider as an ansatz
\begin{equation}
r_k(\vx) = \exp[ \vec{a} \cdot \vx ],
\end{equation}
with $\vec{a}$ to be determined. This ansatz is motivated by the fact that the dominant eigenvalue $0$ for $\mathcal{L}^\dagger$ and $\mathcal{L}$ corresponds to eigenfunctions with an exponential form: \eqref{eq:rho-sfd} for the former, and trivially $e^0$ for the latter.

We observe that $\vec{1} \cdot \hat{\vec{n}}(\vx) = 0$ for all $\vx \in \del \Omega$, and that $\vec{1}$ is the only vector with this property. From the boundary condition \eqref{eq:rk-bc}, we therefore find
\begin{equation}
\tfrac{1}{2}\mathsf{D}(\vec{a} + k \vec{g}) = \alpha \vec{1}
\end{equation}
for some constant $\alpha$. Hence
\begin{equation}
\vec{a} = 2\alpha \mathsf{D}^{-1} \vec{1} - k \vec{g}.
\end{equation}
The periodicity \eqref{eq:periodicity} requires $\vec{a} \cdot \vec{1} = 0$, so that
\begin{equation}
\alpha = \frac{k \vec{1}\cdot \vec{g}}{2\vec{1}^\top \mathsf{D}^{-1} \vec{1}} = \tfrac{1}{2}k \widetilde{D},
\end{equation}
where we have used the property $\vec{1} \cdot \vec{g} = 1$ and \eqref{eq:v-D-tilde}. Applying $\mathcal{L}_k$ to $r_k$, one then finds that $r_k$ is an eigenfunction with eigenvalue
\begin{equation}\label{eq:right-lambda-sfd}
\lambda(k) = 2\widetilde{D}^{-1} \alpha ( \alpha + \tilde{v} ) = k \tilde{v} + \tfrac{1}{2}k^2 \widetilde{D}.
\end{equation}
By LF transform, we then obtain the rate function
\begin{equation}\label{eq:sfd-rf}
I(v) = \frac{(v- \tilde{v})^2}{2 \widetilde{D}},
\end{equation}
which shows that the fluctuations of the current are Gaussian around the stationary velocity $\tilde{v}$.

The same eigenvalue \eqref{eq:right-lambda-sfd} is obtained by assuming for the left eigenfunction 
\begin{equation}
\ell_k = \exp[\vec{b} \cdot \vx ],
\end{equation}
for which we find, in a calculation analogous to the one for $r_k$,
\begin{equation}
\vec{b} =  2\mathsf{D}^{-1}(\vec{v} + \beta \vec{1}) + k \vec{g}
\end{equation}
with
\begin{equation}
\beta = - \frac{\vec{1}^\top \mathsf{D}^{-1} \vec{v} + \tfrac{1}{2} k \vec{1} \cdot \vec{g}}{\vec{1}^\top \mathsf{D}^{-1} \vec{1}} = - \tilde{v} - \tfrac{1}{2} k \widetilde{D}.
\end{equation}

It is interesting to note that the rate function \eqref{eq:sfd-rf} is equivalent to that of a single particle with drift $\tilde{v}$ and diffusivity $\widetilde{D}$. To better understand how current fluctuations are created, we can determine the effective drift $\vec{F}_k$ of the conditioned dynamics using \eqref{eq:Fk} and the expression for $r_k$. The result is
\begin{equation}
\vec{F}_k = \vec{v} + (v - \tilde{v})\vec{1},
\end{equation}
which gives for the probability current
\begin{equation}
\vec{J}_{\vec{F}_k, \rho_k^* }(\vx) = \frac{v}{\tilde{v}} \vec{J}_{\vec{F},\rho^*}(\vx).
\end{equation}
Thus, for the particle system to generate an atypical fluctuation of the collective current, each particle gives rise through fluctuation to an equal absolute increase in their intrinsic velocity, equal to $\Delta v = v - \tilde{v}$. The current thus changes uniformly. 

These results are consistent with the fact that the rate function saturates a universal quadratic bound on current fluctuations \cite{Gingrich2016}, and are also expected given that the heterogeneous single-file diffusion on a ring, while not satisfying detailed balance directly, does so with respect to a reference frame moving with the collective velocity $\tilde{v}$. As a result, the stationary density of the effective process \eqref{eq:rhok*} must be equal to the original invariant density \eqref{eq:rho-sfd}:
\begin{equation}
\rho_k^*(\vx) = \rho^*(\vx).
\end{equation}
This can be checked more directly by noting that $\vec{a}+\vec{b} = \vec{k}$.

To close, it is instructive to compare the results of the single-file diffusion model to those of the asymmetric simple exclusion process (ASEP) conditioned on a large current \cite{Popkov2010}. For that model, it was found analytically that asymptotically large currents are generated through an effective process comprised of two effects: a uniform increase in the hopping rate of all particles, and a pairwise repulsive interaction between particles, not present in the unconditioned process. Since the ASEP yields in the diffusive limit a single-file dynamics with identical particles, we conclude that the second effect is purely a lattice effect. Indeed, on the lattice, jammed configurations form a finite fraction of all possible system configuration, whereas on the continuum, jammed configurations constitute a boundary layer of measure zero relative to the bulk. 

\section{Discussion}

We have provided two independent methods showing that for reflected diffusions, the correct boundary conditions for the spectral problem associated with the dynamical large deviations of current-like observables are given by \eqref{eq:lk-rk-bc}. These boundary conditions are interesting in that they mimic the ``tilting’’ of $\mathcal L$ and $\mathcal{L}^\dagger$ to $\mathcal{L}_k$ and $\mathcal{L}_k^\dagger$, respectively. We indeed recall that $\mathcal{L}_k$ is obtained from $\mathcal{L}$ by the replacement $\nabla \to \nabla + k \vec g$, while  $\mathcal{L}_k^\dagger$ is obtained from  $\mathcal{L}^\dagger$ by the replacement  $-\nabla \to - \nabla + k \vec g$. The same replacements, when applied to the condition \eqref{eq:f-bc} for $f$ and the condition \eqref{eq:reflection-bc} for $\rho$, gives, respectively, the boundary condition \eqref{eq:rk-bc} for $r_k$ and the boundary condition \eqref{eq:lk-bc} for $\ell_k$. Based on this result, it is natural to conjecture that the same replacements apply to other boundary conditions describing other types of boundary behavior, e.g., partially reflecting \cite{singer2008} or sticky \cite{engelbert2014}.

%
%

Two physically significant results follow from the boundary conditions \eqref{eq:lk-rk-bc}. Firstly, they imply that the effective process, describing how current fluctuations are realized, also has reflective boundaries, but relative to the effective drift $\vec F_k$ \eqref{eq:Fk}. This is expected: all system trajectories are reflected at the boundary and, therefore, so is any subset of trajectories corresponding to a given current fluctuation. The second, less obvious result is that the effective force at the boundary is not modified in the normal direction, as expressed in \eqref{eq:Fk=F}. A physical (as opposed to mathematical) understanding of why this is so may come from studying more specific model diffusions. Note that both results were also found for occupation-like observables in one-dimensional diffusions \cite{Buisson2020}, so the type of observable considered (occupation-like or current-like) is not relevant for their explanation. 

In our example of heterogeneous single-file diffusion conditioned on the collective particle current, the effective force changes in both magnitude and direction, but its projection onto the boundary normal does not. One can note that the original process has an irreversible drift \cite{Graham1971}, generally defined by $\vec{F}^\text{ir}(\vec{x}) = \vec{J}^*(\vec{x}) / \rho^*(\vec x)$, which is a constant vector, everywhere orthogonal to the boundary normal. This property allows us to solve the model exactly \cite{Mallmin2021a}, and explains why the irreversible drift of the effective process is only modified in magnitude \cite{Nardini2018}. To obtain more complicated and interesting behavior upon conditioning on a current, one could study models for which the irreversible drift does not have this special structure; looking, for instance, at non-planar boundaries and state-dependent original drifts. Our general results show how, in principle, one can calculate the large deviation elements in these cases. In any such model an interesting question will be the relative importance of the system bulk to the near-boundary region in generating fluctuations.

\appendix

\section{Duality for Markov operators}
\label{app:duality}

We show here the calculation leading to the duality relation for the Markov operators:
\begin{equation}\label{eq:A1}
    \avg{\rho, \mathcal{L} f} = \avg{\mathcal{L}^{\dagger} \rho, f} - \int_{\po}\dx f(\vx) \vec{J}_{\vec{F}, \rho}(\vx) \cdot \hat{\vec{n}}(\vx)  - \frac{1}{2} \int_{\po} \dx \rho(\vx) \mathsf{D} \nabla f(\vx) \cdot \hat{\vec{n}}(\vx) ,
\end{equation}
when the process is constrained to a region $\om \subset \mathbb{R}^d$. Before proceeding, we state for convenience a mathematical identity, which amounts to integration by parts in higher dimensions. For a scalar field $u$ and vector field $\vec{V}$ we have 
\begin{equation} \label{eq:int-by-parts}
    \int_{\Omega} \dx u(\vx) \nabla \cdot \vec{V}(\vx) = -\int_{\partial \Omega}\dx u(\vx) \vec{V}(\vx)\cdot \hat{\vec{n}}(\vx)  - \int_{\Omega}\dx \vec{V}(\vx) \cdot \nabla u(\vx) ,
\end{equation}
where $\hat{\vec{n}}(\vx)$ is the inward normal vector at point $\vx \in \po$. Starting from the inner product over the domain $\om$ and using the expression \eqref{eq:L} for the Markov generator, we write  
\begin{equation} \label{eq:appendA0}
    \avg{\rho, \mathcal{L} f} = \int_{\om} \dx\rho(\vx) \left[\vec{F}(\vx) \cdot \nabla + \frac{1}{2} \nabla \cdot \mathsf{D} \nabla \right] f(\vx) .
\end{equation}
Using \eqref{eq:int-by-parts}, we have 
\begin{equation} \label{eq:appendA1}
    \int_{\om} \dx\rho(\vx) \vec{F}(\vx) \cdot \nabla f(\vx)  = - \int_{\po}\dx \rho(\vx) f(\vx) \vec{F}(\vx) \cdot \hat{\vec{n}}(\vx)   - \int_{\om}\dx \nabla \cdot \left[\vec{F}(\vx) \rho(\vx) \right] f(\vx) 
\end{equation}
and 
\begin{equation} \label{eq:appendA2}
    \int_{\om} \dx \rho(\vx) \left[\frac{1}{2} \nabla \cdot \mathsf{D} \nabla \right] f(\vx)  = - \frac{1}{2}\int_{\po} \dx  \rho(\vx) \mathsf{D} \nabla f(\vx) \cdot \hat{\vec{n}}(\vx)  - \frac{1}{2} \int_{\om} \dx \nabla \rho(\vx) \cdot \mathsf{D} \nabla f(\vx).
\end{equation}
Given that $\mathsf{D}$ is symmetric, we have 
\begin{equation}
    \frac{1}{2}\int_{\om} \dx \nabla \rho(\vx) \cdot \mathsf{D} \nabla f(\vx) = \frac{1}{2}\int_{\om} \dx \mathsf{D}\nabla \rho(\vx) \cdot  \nabla f(\vx) 
\end{equation}
and applying \eqref{eq:int-by-parts} to this last expression, we obtain 
\begin{equation} \label{eq:appendA3}
    \frac{1}{2}\int_{\om} \dx \mathsf{D}\nabla \rho(\vx) \cdot  \nabla f(\vx)  = -\frac{1}{2} \int_{\po}\dx f(\vx) \mathsf{D} \nabla \rho(\vx) \cdot \hat{\vec{n}}(\vx)   - \int_{\om}\dx f(\vx) \left[\frac{1}{2} \nabla \cdot \mathsf{D} \nabla \right]\rho(\vx) . 
\end{equation}
Substituting \eqref{eq:appendA1}, \eqref{eq:appendA2} and \eqref{eq:appendA3} into \eqref{eq:appendA0}, we obtain 
\begin{align}
    \avg{\rho, \mathcal{L} f} =
    \begin{multlined}[t]
         \avg{\mathcal{L}^{\dagger} \rho, f} - \int_{\po}\dx f(\vx) \left[\rho(\vx) \vec{F}(\vx) - \frac{1}{2} \mathsf{D} \nabla \rho(\vx) \right]\cdot \hat{\vec{n}}(\vx)  \\ - \frac{1}{2} \int_{\po} \dx \rho(\vx) \mathsf{D} \nabla f(\vx) \cdot \hat{\vec{n}}(\vx) ,
      \end{multlined}
\end{align}
where $\mathcal{L}^{\dagger}$ is defined as in \eqref{eq:Ldag}. Recognizing the definition of the current \eqref{eq:J} in the above, \eqref{eq:A1} follows.
 
\section{Duality for tilted generators} 
\label{app:duality2}

Here we obtain the duality expression  
\begin{equation}\label{eq:B1}
\avg{\ell_k, \mathcal{L}_k r_k}  = \avg{ \mathcal{L}^\dagger_k \ell_k, r_k} -\int_{\del \Omega}\dx \vec{J}_{\vec{F}_k, \ell_k r_k}(\vx) \cdot \hat{\vec{n}}(\vx) ,
\end{equation}
for the tilted generators for a process constrained to a region $\Omega \subset \mathbb{R}^d$, proceeding in a similar manner as done in \ref{app:duality}. Using the expression \eqref{eq:Lk} for the tilted generator, we have 
\begin{equation}
    \avg{\ell_k, \mathcal{L}_k r_k} = \int_{\om} \dx \ell_k(\vx) \left[\vec{F}(\vx)\cdot (\nabla + k \vec{g}(\vx)) + \frac{1}{2} (\nabla + k \vec{g}(\vx)) \cdot \mathsf{D} (\nabla + k \vec{g}(\vx))  \right] r_k(\vx).
\end{equation}
For the first term we have, using integration by parts as in \eqref{eq:int-by-parts}, that 
\begin{align} \label{eq:appendB2}
    \int_{\om} \dx \ell_k(\vx) \bigg[\vec{F}(\vx)\cdot (\nabla + k \vec{g}(\vx))\bigg] r_k(\vx) =
    \begin{multlined}[t]
    - \int_{\po} \dx l_k(\vx)r_k(\vx) \vec{F}(\vx)  \cdot \hat{\vec{n}}(\vx)  \\ + \int_{\om}\dx  \bigg[(-\nabla + k \vec{g}(\vx)) \cdot \vec{F}(\vx) l_k(\vx) \bigg] r_k(\vx).
    \end{multlined}
\end{align}
For the second term, we first note that 
\begin{equation} \label{eq:appendB1}
    \frac{1}{2} (\nabla + k \vec{g}(\vx)) \cdot \mathsf{D} (\nabla + k \vec{g}(\vx)) = \frac{1}{2} (\nabla \cdot \mathsf{D} \nabla + k \vec{g}(\vx) \cdot \mathsf{D} \nabla + \nabla \cdot k \mathsf{D} \vec{g}(\vx) + k^2 \vec{g}(\vx)\cdot \mathsf{D} \vec{g}(\vx)).
\end{equation}
The last term in the above contains no derivatives and produces no boundary terms, while the first term has already been dealt with in \ref{app:duality} in \eqref{eq:appendA2} and \eqref{eq:appendA3} (with the understanding that $\ell_k$ and $r_k$ are to replace $\rho$ and $f$, respectively). We can therefore immediately write 
\begin{align} \label{eq:appendB3}
    \int_{\om} \dx \ell_k(\vx) \left[\frac{1}{2} \nabla \cdot D \nabla  \right]r_k(\vx) =
    \begin{multlined}[t]
         \int_{\om} \dx r_k(\vx) \left[\frac{1}{2} \nabla \cdot D \nabla  \right]\ell_k(\vx) \\ - \frac{1}{2} \int_{\po}\dx \ell_k(\vx) \mathsf{D}\nabla r_k(\vx) \cdot \hat{\vec{n}}(\vx) \\+ \frac{1}{2} \int_{\po}\dx r_k(\vx) \mathsf{D}\nabla \ell_k(\vx) \cdot \hat{\vec{n}}(\vx).
    \end{multlined}    
\end{align}
For the last two terms in \eqref{eq:appendB1}, we have 
\begin{subequations} \label{eq:appendB4}
\begin{align} 
    \int_{\om} \dx \ell_k(\vx) \left[\frac{1}{2} k \vec{g}(\vx) \cdot \mathsf{D} \nabla \right]r_k(\vx) &=
    \begin{multlined}[t] 
     \int_{\om} \dx \ell_k(\vx) \left[\frac{1}{2} k \mathsf{D} \vec{g}(\vx) \cdot  \nabla \right]r_k(\vx)
     \end{multlined} \\
    &= \begin{multlined}[t]
        -\frac{1}{2} \int_{\po} \dx \ell_k(\vx) r_k(\vx) k \mathsf{D} \vec{g}(\vx) \cdot \hat{\vec{n}}(\vx)\\ - \frac{1}{2}\int_{\om} \dx r_k(\vx) \nabla \cdot \ell_k(\vx) k \mathsf{D} \vec{g}(\vx)
    \end{multlined}    
\end{align}
\end{subequations}
where we have used the symmetry of $\mathsf{D}$ in the first line and integration by parts in the second, and 
\begin{align} \label{eq:appendB5}
    \int_{\om} \dx \ell_k(\vx) \left[\frac{1}{2} \nabla \cdot k \mathsf{D} \vec{g}(\vx) \right]r_k(\vx) = 
    \begin{multlined}[t]
     -\frac{1}{2}\int_{\po} \dx \ell_k(\vx)r_k(\vx) k \mathsf{D} \vec{g}(\vx) \\  - \int_{\om} \dx \left[\frac{1}{2} k \mathsf{D} \vec{g}(\vx) \cdot \nabla \ell_k(\vx) \right] r_k(\vx).
     \end{multlined}
\end{align}
Combining \eqref{eq:appendB2}, \eqref{eq:appendB3}, \eqref{eq:appendB4} and \eqref{eq:appendB5} we obtain
\begin{align}
    \avg{\ell_k, \mathcal{L}_k r_k} =
    \begin{multlined}[t]
         \int_{\om} \dx r_k(\vx) \bigg[(-\nabla + k \vec{g}(\vx))\cdot (\vec{F}(\vx) \ell_k(\vx)) + \frac{1}{2} \bigg(\nabla \cdot \mathsf{D} \nabla \ell_k(\vx)  \\  - k\mathsf{D} \vec{g}(\vx) \cdot \nabla \ell_k(\vx) - \nabla \cdot (k\mathsf{D} \vec{g}(\vx) \ell_k(\vx)) + k^2 \vec{g}(\vx) \cdot \mathsf{D} \vec{g}(\vx) \ell_k(\vx)  \bigg)  \bigg]  \\ 
    	 - \int_{\po} \dx \bigg\{\ell_k(\vx) r_k(\vx)\bigg(\vec{F}(\vx) + k \mathsf{D} \vec{g}(\vx) \bigg) + \frac{1}{2} \ell_k(\vx) \mathsf{D} \nabla r_k(\vx) \\ - \frac{1}{2} r_k(\vx) \mathsf{D} \nabla \ell_k(\vx) \bigg\}\cdot \hat{\vec{n}}(\vx).
    \end{multlined}
\end{align}
Noting from \eqref{eq:Lkdag} that the expression in square brackets is simply the operator $\mathcal{L}_k^{\dagger}$ applied to $\ell_k$, and observing that 
\begin{equation}
    \ell_k r_k\bigg(\vec{F} + k \mathsf{D} \vec{g} \bigg) + \frac{1}{2} \ell_k \mathsf{D} \nabla r_k - \frac{1}{2} r_k \mathsf{D} \nabla \ell_k = 
    \vec{J}_{\vec{F}_k,\ell_k r_k}
\end{equation}
where we have used the definition of the current \eqref{eq:J} and effective force \eqref{eq:Fk}, we obtain the duality relation \eqref{eq:B1}.
 
 \section*{Acknowledgments}
Emil Mallmin acknowledges studentship funding from EPSRC Grant No. EP/N509644/1.
 
\section*{References}

\bibliography{merged}
\bibliographystyle{physicsbibstyle}	

\end{document}